# What artificial intelligence might teach us about the origin of human language

Kilpatrick, Alexander
Nagoya University of Business and Commerce

*This study has not undergone peer review. Proceed with scepticism.*

**ABSTRACT**

This study explores an interesting pattern emerging from research that combines artificial intelligence with sound symbolism. In these studies, supervised machine learning algorithms are trained to classify samples based on the sounds of referent names. Machine learning algorithms are efficient learners of sound symbolism, but they tend to bias one category over the other. The pattern is this: when a category arguably represents greater threat, the algorithms tend to overpredict to that category. A hypothesis, framed by error management theory, is presented that proposes that this may be evidence of an adaptation to preference cautious behaviour. This hypothesis is tested by constructing extreme gradient boosted (XGBoost) models using the sounds that make up the names of Chinese, Japanese and Korean Pokémon and observing classification error distribution.

## 1. INTRODUCTION

For the last hundred years, theoretical approaches for the evolution of human language have been overshadowed by the *gesture first* hypothesis of language origin [1] and the philosophy that the relationship between sound and meaning is arbitrary [2]. On the latter, a growing collection of research has revealed systematic sound symbolic relationships in human language. The conflict between sound symbolism and arbitrariness extends at least as far back as Plato's dialogue *Cratylus* which details a debate between philosophers in which Socrates eventually concludes that sometimes the relationship between form and meaning is arbitrary, and sometimes it is not [3]. Modern research has shown that Socrates is right in this regard, sound symbolism in human language is stochastic, not prescriptive; that is, there is a random probability distribution that can be analysed statistically but may not apply in all instances [4]. Interestingly, many systematic sound symbolic relationships have been found to hold across languages. For example, one study [5] used nonce word stimuli to explore the judgement of size associated with vowels, targeting speakers of Chinese, English, Japanese, and Korean. They found that in all languages, [i] was associated with smaller objects while [a] was associated with larger objects. This pattern has typically been framed by the frequency code hypothesis [6] which suggests that size is reflected by the acoustic properties of sounds. However, some cross-linguistic sound symbolic relationships arguably do not reflect size, but rather shape. For example, a recent study found that the *kiki/bouba* effect—the association of the nonce words like *kiki* with spiky shapes and nonce words like *bouba* with rounded shapes—was robust in 17 of the 25 tested languages [7].

Another cross-linguistic sound symbolic pattern that is arguably reflective of neither size nor shape is found in Pokémonastics. Pokémonastics is the analysis of sound symbolic patterns in the names of video game characters known as Pokémon. Pokémon has been translated into many languages and Pokémon have attributes that remain consistent across languages. An interesting observation by Shih et al. [8] is that sound symbolic effects in Pokémon names are stronger in those traits that are important for in-game survivability. For example, while both combat attributes and gender are shown to exhibit sound symbolic patterns, those patterns are stronger in combat attributes than they are in gender which has little in-game effect. Another attribute that Pokémon possess is that of friendship, which reflects how naturally friendly each Pokémon is. In an examination of the relationship between friendship and plosives in Chinese, English, French, German, Japanese, and Korean Pokémon names [9], the number of times [p] occurs in each name was a significant predictor of high friendship in all languages, while the number of times [g] occurs in each name was a significant predictor of low friendship in all languages that have phonological opposition for voicing on plosives (i.e., not Chinese or Korean). The cross-linguistic nature of these patterns and the apparent lack of size and shape of the dependent variable led the authors to tentatively conclude that this may be evidence for a shared sound symbolism system that may be the result of a somewhat vestigial alarm call system. This is supported by the convergent evolution effect—that different species undergoing similar environmental pressure will evolve similar traits—and the prevalence of alarm call systems in the animal kingdom. For example, non-human primates in sub-Saharan Africa—specifically, the vervet monkey [10] and the green monkey [11]—have complex alarm call systems that communicate conspecifics of predators.

Recent Pokémonastic studies have incorporated supervised machine learning algorithms to further examine patterns in the Pokémon corpus and test the learning efficiency of these models. Using the random forest (RF) algorithm, models were constructed to classify Pokémon into pre- and post-evolution categories based on the sounds that make up their names in Chinese, Japanese, and Korean [12]. Evolution in the Pokémon franchise is a way that the characters "level up". When Pokémon evolve, their names and appearances change, and they become stronger. The initial models were found to be overfitting, so multiple independent RFs were constructed with individually randomised subsets to cross-validate the dataset. All RFs achieved a classification accuracy greater than chance, but an interesting pattern emerged in error distribution and feature importance calculations. The error rates tended to be greater for pre-evolution Pokémon and the features that were found to be important in classification accuracy, tended to be overrepresented in the post-evolution category. Table 1 presents a confusion matrix for the Chinese RF which shows that the model tended to overclassify to the post-evolution category.

|        |               | Classification |                |
|--------|---------------|----------------|----------------|
|        |               | Pre-evolution  | Post-evolution |
| Sample | Pre-evolution | 479            | 444            |
|        | Post-evolution| 300            | 667            |

**Table 1**: Confusion matrix for the Chinese Pokémon random forest algorithm [12].

|        |        | Classification |       |
|--------|--------|----------------|-------|
|        |        | Female         | Male  |
| Sample | Female | 9539           | 12623 |
|        | Male   | 6461           | 28918 |

**Table 2**: Confusion matrix for the Chinese given name random forest algorithm [13].

This pattern has also been observed in similar experiments [13,14] that construct RFs on the sounds that make up the given names of humans. These RFs are designed to explore the sound symbolism of gender in given names in both Chinese [13] and Japanese [14]. A confusion matrix for an RF constructed using Chinese given names is presented in Table 2.

Although [12] and [13] test two-tailed hypotheses, interpreting those findings through the lens of a one-tailed hypothesis in the present study might offer new insight into sound symbolism. In a one-tailed test, post-evolution in Table 1 and the male gender in Table 2 might be considered as the more threatening (and therefore more likely to elicit an alarm call) of each classification category. Those samples from the non-threatening categories that were classified to the threatening categories would then be false positive (FP) errors and those threatening samples that were classified as non-threatening would be false negative (FN) errors. Through this interpretation, FP error rate can be calculated by adjusting percentage of FP errors by distribution skew in the testing subset. In table 1 for example, there is a 4.8% greater number of post-evolution samples (967) than pre-evolution samples (923), FP error rate is the percentage of FP errors with the minority error multiplied by 104.8% to account for skew. FP error rate equals 58% for the algorithms presented in Table 1 and 76% for the algorithms presented in Table 2. Note here that these are aggregated results and that a minority of RFs do not exhibit an FP error rate greater than 50%. Interestingly, distribution skew of important features tends to be towards threatening categories. In other words, those features that are important to the models in making classification decisions have a greater distribution to the post-evolution and male categories. For example, the most important features for the Chinese Pokémon RF presented in Table 1 were the falling tone, /ŋ/, /t͡ɕ/, /ɕ/, and /e/. Of these, only /ɕ/ had a greater distribution to the pre-evolution category and only the falling tone and /ŋ/—which were overrepresented in post-evolution samples—returned a significant feature importance. This pattern is typical of all the models presented in Table 3 which reports model accuracy and FP error rate for five multiple RFs. These are just a sample; more examples of this pattern can be found in [12].

| Dataset              | Dependent Var. | Accuracy. | FP% | Ref. |
|----------------------|----------------|-----------|-----|------|
| Chinese Pokémon      | Evolution      | 60%       | 58% | [12] |
| Chinese Given Names  | Gender         | 72%       | 76% | [13] |
| Japanese Pokémon     | Evolution      | 68%       | 52% | [12] |
| Japanese Given Names | Gender         | 76%       | 54% | [14] |
| Korean Pokémon       | Evolution      | 60%       | 60% | [12] |

**Table 3**: Model accuracy and FP error rate (FP%) for the five multiple RFs.

An error distribution skew that preferences FP errors in the communication of threats is potentially evolutionarily advantageous. *Error management theory* [15], based on *signal detection theory*, is a framework for evaluating how organisms make decisions amidst uncertainty. EMT posits that evolutionary selection favours organisms that opt for behaviour that would result in the least costly error. For example, vervet monkeys give acoustically different alarm calls to leopards, eagles, and snakes [10]. Infant vervet monkeys seem to be predisposed from birth to broadly divide other species into predators and non-predators, but tend to make mistakes, giving alarm calls to non-predators such as pigeons and warthogs [16]. An infant vervet monkey giving an eagle alarm for a pigeon is an example of an FP error and would be potentially less costly than if the infant were to make a FN error and fail to signal the presence of a predator. Through the lens of the hypothesis that certain elements of sound symbolism in human language communicate threat, EMT is a potential explanation for the overrepresentation of FP errors observed in machine learning algorithms.

## 2. METHOD AND MATERIALS

In this study, XGBoost models [17] are constructed using the names of Japanese, Korean, and Chinese Pokémon taken from an online repository [18]. This study uses XGBoost models, and not RF models, to test if error distribution skew occurs outside of the RF algorithm. The names are converted into a count of the number of times each speech feature occurs in each name. This results in a dataset of 898 samples with features comprised mostly of null values which has been found to cause overfitting in decision tree-based algorithms [12]. To correct for this, k-fold cross-validation (k = 3) is used which means that three iterations of each model is constructed. Each iteration uses different subsets for the training (A+B, A+C, and B+C) and testing (C, B, and A) of the algorithms. The reported results are taken from the aggregation of these three iterations, but the statistical hypothesis testing is conducted on the results of individual iterations. For each language, four dependent variables are tested, these are Attack (a measure of how powerful each Pokémon's attacks are), Defend (a measure of how well each Pokémon can defend themselves), Height, and Weight. These variables were chosen because an increase value for each should theoretically represent an increased threat, but that increase should be different among variables. For example, an increase in those parameters that are combat specific (Attack and Defend) should represent a greater increase in threat than those parameters that are size specific (Height and Weight). Important to note that in the video games for which the names were devised, Attack and Defend have direct effects on combat, while Height and Weight have no in game effect. Sample classification is derived from a median split of each continuous variable with median samples being omitted from the dataset and categories being balanced by randomly omitting samples from the majority category for each dependent variable. This study tests the following hypotheses; **H1**: FP error will be greater than 50% in all models; **H2**: Models classifying combat parameters will have higher FP error than size specific models. Data are available at the following URL: https://tinyurl.com/mw9u43ya

## 3. RESULTS

All Machine learning algorithms achieved an accuracy of greater than 50%. FP for combat models was greater than 50%, but this was not the case for all size models. To test **H1**, intercept only linear models were calculated against 50% chance. To test **H2**, simple linear models were conducted. All models used the data from individual iterations (36 total algorithms), not the aggregated results presented in Table 4.

|  | Attack | | Defend | |
|---|---|---|---|---|
|  | Accuracy | FP% | Accuracy | FP% |
| Japanese | 59.15% | 51.93% | 59.16% | 55.96% |
| Chinese | 57.04% | 57.08% | 55.66% | 53.35% |
| Korean | 56.10% | 55.47% | 54.07% | 55.64% |
| Average | 57.43% | 54.83% | 56.30% | 54.98% |
|  | Height | | Weight | |
|  | Accuracy | FP% | Accuracy | FP% |
| Japanese | 63.03% | 48.82% | 62.33% | 47.97% |
| Chinese | 59.27% | 44.17% | 57.51% | 54.05% |
| Korean | 57.77% | 48.77% | 55.38% | 51.50% |
| Average | 60.03% | 47.26% | 58.41% | 51.17% |

**Table 4**: Aggregated accuracy and FP error rate for the XGBoost models.

**H1)** Of the four dependent variables, only the Attack model revealed a significant interaction on FP error distribution; $t(8) = 2.6, p = 0.031$. Defend ($p = 0.145$), Height ($p = 0.522$), and Weight ($p = 0.699$) models were not significant. Intercept only models calculated by combining combat and size variables showed that only combat variables were significant; $t(17) = 2.8, p = 0.012$, while the size variables were not ($p = 0.757$). **H2)** A simple linear model showed that FP in combat variables ($M = 55\%, SD = 7\%$) was significantly different from FP in size variables ($M = 49\%, SD = 10\%$); $t(34)=25.55, p < 0.001$.

One potential explanation for increased FP in combat models is the *longer-is-stronger* principle [19] which is the finding that stronger Pokémon generally have longer names. To explore its potential effects, a series of simple linear regression models were conducted to examine the relationship between the four continuous variables (including previously omitted samples) and a count of the number of features in names excluding tones in Chinese. All linear regression analyses revealed a significant positive correlation ($p < 0.05$) except for Height ($p = 0.67$) and Weight ($p = 0.70$) in the Chinese dataset. The results of a combined analysis are presented in Table 5.

| Variable | Deg. Freedom | F | p | $R^2$ |
|---|---|---|---|---|
| Attack | 1, 2693 | 58.88 | 2.33E-14 | 0.021 |
| Defend | 1, 2693 | 49.36 | 2.69E-12 | 0.018 |
| Height | 1, 2693 | 17.95 | 2.35E-05 | 0.007 |
| Weight | 1, 2693 | 27.40 | 1.79E-07 | 0.010 |

**Table 5**: Simple linear regression analysis of relationships between parameters and name length.

## 4. DISCUSSION

Contrary to H1, not all dependent variables had an FP error greater than 50%. Only the Attack model was significantly different from chance; although when combined with Defend, the combat model was significant. Height and Weight have no in game effect which might explain why size models did not follow the pattern of combat models. In alignment with H2, combat model FP error rates were significantly higher than size model FP error rates. Despite using a different machine learning algorithm, those patterns exhibited by the combat models in this study conform to the models presented in Table 3. These findings are in alignment with the central hypothesis that there would be an overrepresentation of FP errors in algorithms that classify according to threat. In the context of an alarm call system, this would mean that ambiguous vocalisations would be more likely to be perceived as alarm calls rather than not. Although, this would result in more FP errors, those errors are less costly than FN errors, hence the strategy is evolutionarily beneficial. Any evolutionary adaptation based on vocalisation and present in modern language would suggest that human language did not evolve entirely in the kinaesthetic modality as promulgated by the gesture first hypothesis which has recently come under criticism for this position [20].

Although it is not entirely clear from the models themselves, the most likely explanation for an FP error skew is that those features that are important in the classification of threat are more likely to occur in non-threatening samples than those features that are important in the classification of non-threats are to occur in threatening samples. This effect would be difficult to observe in more traditional statistical analyses because, unless it was being tested for specifically, it would likely manifest as reduced effect size across the model, depending on the test. The cross-validation methods in [12], [13], [14], and the present study have a secondary function being that they produce multiple results which can be analysed with traditional methods to test for significant distribution skew.

A potential alternative explanation for this is the *longer is stronger* principle. Longer names have fewer null values. Decision tree-based algorithms construct many decision trees—made from randomly bagged samples [21]. Each tree consists of nodes and each node tests a random selection of features [22]. If the node selects only those features with null values, then the node has no function, and the outcome at the terminal node must be based on alternate nodes. There are three issues with this explanation: firstly, it is unclear how this might cause the algorithm to overclassify to the threat category, second, the hyperparameter governing the number of features selected at each node is tuned to the dataset resulting in fewer functionless nodes, and lastly, the linear regression analysis showed a significant positive correlation between all variables and name length except in Chinese. Although the effect of length is stronger in combat parameters, one would reasonably assume that if this explanation were accurate, increased name length should increase FP error for all variables—not just combat variables.

The *Sagan standard* posits that extraordinary claims require extraordinary evidence. The present study uses a limited corpus of video game character names, not natural language. It only examines East Asian languages. Much like the RF algorithm, it is difficult to observe the inner workings of the XGBoost algorithm; therefore, it is not entirely clear why the algorithms are preferencing FP errors. Probably the most salient issue with the present study—and its central hypothesis—is that it anthropomorphises machine learning algorithms, i.e., it assumes that those patterns that cause the artificial intelligence to increase FP error should have the same effect in humans. It is the recommendation of the author that more research in this area be conducted prior to the development of any concrete conclusions.

## 4. CONCLUSION

This study presents XGBoost model algorithms that are designed to classify Pokémon names into high or low Attack, Defend, Height, and Weight categories based on sound symbolism. It tests whether these algorithms exhibit a distribution skew towards higher categories which are argued to be representative of increased threat. Only those models that classify combat variables are found to have significantly increased FP error distribution. This is potential evidence that human language has an inbuilt feature that means that sound symbolic information is perceived as threat more often than not. Through the lens of EMT, this is evolutionarily beneficial because FP errors are less costly than FN errors. Although more research is recommended, this finding has the potential to significantly impact our understanding of sound symbolism and the origin of human language.